\documentclass[journal=jctcce,manuscript=article]{achemso}
\usepackage{color}
\usepackage{array}
\usepackage{mathtools}
\usepackage{multirow}
\usepackage{amsmath}
\usepackage{amssymb}
\usepackage{graphicx}

\makeatletter

\providecommand{\tabularnewline}{\\}

\usepackage{chemformula}
\usepackage[version=3]{mhchem}
\usepackage{xcolor}
\usepackage{longtable}
\usepackage{tablefootnote}
\usepackage{tcolorbox}
\usepackage{enumerate}
\usepackage{multirow}
\usepackage{subcaption}

\newcommand{\onlinecite}[1]{\hspace{-2 ex} \citenum{#1}}

\author{Chaoqun Zhang}
\email{czhan119@jhu.edu}
\affiliation{Department of Chemistry, The Johns Hopkins University, Baltimore, MD 21218, USA}
\author{Filippo Lipparini}
\affiliation{Dipartimento di Chimica e Chimica Industriale, Universit\`a di Pisa, Via G. Moruzzi 13, I-56124 Pisa, Italy}
\author{Stella Stopkowicz}
\affiliation{Fachrichtung Chemie, Universität des Saarlandes, D-66123 Saarbrücken, Germany}
\author{Jürgen Gauss}
\affiliation{Department Chemie, Johannes Gutenberg-Universität Mainz, Duesbergweg 10-14, D-55128 Mainz, Germany}
\author{Lan Cheng}
\affiliation{Department of Chemistry, The Johns Hopkins University, Baltimore, MD 21218, USA}

\title
  {A Cholesky decomposition-based implementation of relativistic two-component coupled-cluster methods for medium-sized molecules}

\makeatother

\begin{document}

\begin{abstract}
A Cholesky decomposition (CD)-based implementation of relativistic two-component coupled-cluster (CC) and equation-of-motion CC (EOM-CC) methods using an exact two-component Hamiltonian augmented with atomic-mean-field spin-orbit integrals (the X2CAMF scheme) is reported. 
The present CD-based implementation of X2CAMF-CC and EOM-CC methods employs atomic-orbital-based algorithms to avoid the construction of two-electron integrals and intermediates involving three and four virtual indices.
Our CD-based implementation extends the applicability of X2CAMF-CC and EOM-CC methods to medium-sized molecules with the possibility to correlate around 1000 spinors.
Benchmark calculations for uranium-containing small molecules have been performed to assess the dependence of the CC results on the Cholesky threshold. 
A Cholesky threshold of $10^{-4}$ is shown to be sufficient to maintain chemical accuracy.
Example calculations to illustrate the capability of the CD-based relativistic CC methods are reported for the bond-dissociation energy of the uranium hexafluoride molecule, \ce{UF6}, with up to quadruple-zeta basis sets, and 
the lowest excitation energy in solvated uranyl ion [\ce{UO2^2+(H2O)12}].
\end{abstract}

\section{Introduction}

Molecules containing heavy elements play increasingly important roles
in a variety of research areas, ranging from catalysis, \cite{shibasakiLanthanideComplexesMultifunctional2002,weissOrganofelementCatalystsEfficient2010,arnoldCarbonOxygenateTransformations2017}
medicine, \cite{loehrerCisplatin1984,zhangNewMetalComplexes2003,monroTransitionMetalComplexes2018,luRecentDevelopmentGold2022}
optoelectronic devices,\cite{hasegawaStrategiesDesignLuminescent2004,youPhotofunctionalTripletExcited2012,costaLuminescentIonicTransitionmetal2012}
to quantum simulation, \cite{carrColdUltracoldMolecules2009,yuScalableQuantumComputing2019,saskinNarrowlineCoolingImaging2019,madjarovHighfidelityEntanglementDetection2020}
and the search of new physics beyond the Standard Model.\cite{demilleProbingFrontiersParticle2017,kozyryevPrecisionMeasurementTimeReversal2017,safronovaSearchNewPhysics2018,cairncrossAtomsMoleculesSearch2019}
Accurate treatments of both relativistic effects \cite{pyykkoRelativityPeriodicSystem1979,pyykkoRelativisticEffectsStructural1988,pitzerRelativisticEffectsChemical1979,pitzerElectronicstructureMethodsHeavyatom1988,dyallIntroductionRelativisticQuantum2007a,liuIdeasRelativisticQuantum2010,saueRelativisticHamiltoniansChemistry2011,autschbachPerspectiveRelativisticEffects2012,reiherRelativisticQuantumChemistry2014}
and electron correlation are required for reliable predictions of molecular properties for heavy-element-containing molecules.
Combining coupled-cluster (CC) \cite{shavittMANYBODYMETHODSCHEMISTRY} and equation-of-motion CC (EOM-CC) methods \cite{stantonEquationMotionCoupled1993,krylovEquationofmotionCoupledclusterMethods2008} with relativistic four- and two-component Hamiltonians, \cite{dyallIntroductionRelativisticQuantum2007a} the spinor-based relativistic CC and EOM-CC methods \cite{eliavOpenshellRelativisticCoupledcluster1994,eliavRelativisticCoupledCluster1994a,visscherKramersrestrictedClosedshellCCSD1995,visscherFormulationImplementationRelativistic1996a,leeSpinorbitEffectsCalculated1998a,natarajGeneralImplementationRelativistic2010,visscherFormulationImplementationRelativistic2001,kouliasRelativisticRealTimeTimeDependent2019,liuRelativisticCoupledclusterEquationofmotion2021} and analytic derivative techniques \cite{sheeAnalyticOneelectronProperties2016,liuAnalyticEvaluationEnergy2021,zhengGeometryOptimizationsSpinorbased2022} have emerged as useful tools
for chemical applications involving heavy-element-containing molecules aiming at high accuracy.
Among the relativistic CC methodologies, the four-component CC methods based on the Dirac-Coulomb(-Breit) Hamiltonian \cite{dyallIntroductionRelativisticQuantum2007a,breitEffectRetardationInteraction1929} are the most rigorous but also computationally most expensive. 
The need to handle a large number of relativistic two-electron atomic-orbital (AO) integrals
and large molecular-orbital (MO) integral matrices has so far limited the applicability of four-component CC methods to relatively small molecules.
Relativistic two-component methods \cite{hessRelativisticElectronicstructureCalculations1986,vanlentheRelativisticRegularTwocomponent1996,dyallInterfacingRelativisticNonrelativistic1997,nakajimaNewRelativisticTheory1999,baryszTwocomponentMethodsRelativistic2001,liuExactTwocomponentHamiltonians2009,liuIdeasRelativisticQuantum2010,saueRelativisticHamiltoniansChemistry2011} have been developed to reduce the computational cost by decoupling the positronic and electronic degree of freedoms, among which the exact two-component (X2C) approach \cite{dyallInterfacingRelativisticNonrelativistic1997,kutzelniggQuasirelativisticTheoryEquivalent2005,iliasInfiniteorderTwocomponentRelativistic2007,liuExactTwocomponentHamiltonians2009} appears to be the most promising.
The X2C molecular mean-field (X2CMMF) approach eliminates relativistic AO two-electron integrals in the integral transformation and thus significantly reduces the cost of the integral-transformation step. \cite{sikkemaMolecularMeanfieldApproach2009} X2CMMF-CC calculations can also benefit from AO-based algorithms.\cite{liuTwocomponentRelativisticCoupledcluster2018,asthanaExactTwocomponentEquationofmotion2019}
High-level parallelization using graphic processing units (GPUs) has been recently also reported for X2CMMF-CC calculations 
and can significantly speed up the floating-point operations involved in the calculations \cite{ufimtsevQuantumChemistryGraphical2008,deprinceCoupledClusterTheory2011,pototschnigImplementationRelativisticCoupled2021}.
On the other hand, the dependence of the mean-field spin-orbit integrals on the molecular four-component Hartree-Fock solutions leads to complications in the development of analytic derivative techniques for the X2CMMF scheme.
Specifically, the evaluation of orbital-relaxation contributions to analytic X2CMMF derivatives requires 
including the response of the mean-field spin-orbit integrals.
An analytic evaluation of electrical properties for the spin-free version of X2CMMF-CC
with a solution of coupled-perturbed Hartree-Fock equations to account for the response of mean-field spin-orbit integrals\cite{kirschAnalyticEvaluationFirstorder2019} and an unrelaxed version of analytic X2CMMF-CCSD first derivatives \cite{sheeAnalyticOneelectronProperties2016} have been reported so far. 

To address the existing challenges in the development of analytic X2CMMF gradients
and to further improve computational efficiency without significant loss of accuracy, 
we have developed an exact two-component Hamiltonian with atomic mean-field \cite{hessMeanfieldSpinorbitMethod1996} spin-orbit integrals (the X2CAMF scheme). \cite{liuAtomicMeanfieldSpinorbit2018,zhangAtomicMeanfieldApproach2022}
The X2CAMF scheme has been shown to treat scalar-relativistic effects, spin-orbit coupling, and the Breit interaction with high accuracy and efficiency. 
AO-based algorithms 
have been developed for the X2CAMF-based 
CC singles and doubles (CCSD) \cite{purvisiiiFullCoupledCluster1982} 
and EOM-CC singles and doubles (EOM-CCSD) \cite{stantonEquationMotionCoupled1993} methods
in order to eliminate the computational bottlenecks associated with large MO integral files involving four virtual indices.
Analytic gradients \cite{liuAnalyticEvaluationEnergy2021,zhengGeometryOptimizationsSpinorbased2022} for these methods have been implemented and enable the efficient and accurate calculation of molecular properties including equilibrium geometries and harmonic vibrational frequencies for polyatomic molecules containing heavy atoms.
On the other hand, a further extension of the applicability of X2CAMF-CC methods to medium-sized molecules with around 1000 spinors is required to deal with the rapid increase of the size of AO two-electron integrals and MO integral matrices involving three virtual indices.
The use of techniques to decompose the two-electron integral matrices to exploit their intrinsic sparsity, including density-fitting/resolution-of-identity (DF/RI) 
\cite{whittenCoulombicPotentialEnergy1973,baerendsSelfconsistentMolecularHartree1973,sambeNewComputationalApproach1975,vahtrasIntegralApproximationsLCAOSCF1993,eichkornAuxiliaryBasisSets1995a,weigendRIMP2FirstDerivatives1997,weigendRIMP2OptimizedAuxiliary1998}, Cholesky decomposition (CD) \cite{beebeSimplificationsGenerationTransformation1977,kochReducedScalingElectronic2003,aquilanteQuarticScalingEvaluation2007,pedersenDensityFittingAuxiliary2009}, and
tensor hypercontraction \cite{hohensteinTensorHypercontractionDensity2012,parrishExactTensorHypercontraction2013,zhaoOpenShellTensorHypercontraction2023} methods, 
appears to be a promising step forward. 

Introduced to quantum chemistry in the 1970s, the DF/RI and CD methods have been proven very efficient in reducing the computational cost and storage requirements in non-relativistic quantum-chemical calculations. 
The two schemes are closely related to each other. 
CD can be viewed as an automatic procedure to generate auxiliary basis functions for density fitting by removing the linear dependence of basis functions in the Coulomb metric.\cite{pedersenDensityFittingAuxiliary2009}
It provides controllable errors through a predefined Cholesky threshold, making it compatible with any type of Hamiltonian and suitable for high-accuracy calculations. 
On the other hand, the use of pre-optimized auxiliary basis functions in the DF/RI methods, although method- and Hamiltonian-specific, is generally computationally more efficient. 

The DF/RI and CD techniques have been implemented and are being routinely used together with many quantum-chemical methods, including Hartree-Fock (HF), \cite{vahtrasIntegralApproximationsLCAOSCF1993,weigendFullyDirectRIHF2002}
density-functional theory (DFT), \cite{skylarisResolutionIdentityCoulomb2000,eichkornAuxiliaryBasisSets1995a}
second-order M{\o}ller-Plesset perturbation theory (MP2),\cite{feyereisenUseApproximateIntegrals1993,weigendRIMP2FirstDerivatives1997,weigendRIMP2OptimizedAuxiliary1998,burgerNMRChemicalShift2021,blaschkeffCDMP22022}
random-phase approximation (RPA),\cite{eshuisFastComputationMolecular2010}
coupled-cluster methods,\cite{hattigGeometryOptimizationsCoupledcluster2003,kohnAnalyticGradientsExcited2003,pedersenPolarizabilityOpticalRotation2004,epifanovskyGeneralImplementationResolutionoftheidentity2013}
and multireference approaches. \cite{aquilanteAccurateInitioDensity2008,aquilanteCholeskyDecompositionbasedMulticonfiguration2008,bostromCalibrationCholeskyAuxiliary2010,nottoliSecondorderCASSCFAlgorithm2021,nottoliChemicalShiftCASSCF2022}
CD is also a fundamental component of auxiliary-field quantum Monte-Carlo methods.\cite{honmaDiagonalizationHamiltoniansManyBody1995,mottaInitioComputationsMolecular2018}
The CD-based non-relativistic CC methods, \cite{pedersenPolarizabilityOpticalRotation2004,epifanovskyGeneralImplementationResolutionoftheidentity2013} along with their analytical gradients \cite{fengImplementationAnalyticGradients2019,schnack-petersenEfficientImplementationMolecular2022} effectively reduce storage and input/output (I/O) requirements associated with the
two-electron integrals, intermediates, and two-particle density matrices.
Although DF/RI and CD do not reduce the floating-point operation counts, the reduction in memory and disk-space requirement as well as I/O facilitates calculations of larger molecules. 
The DF/RI and CD have also been used in relativistic quantum-chemical calculations. 
Density-fitted four-component HF and complete active space self-consistent field methods with Dirac-Coulomb-Breit Hamiltonian have been reported \cite{kelleyLargescaleDiracFock2013,batesFullyRelativisticComplete2015},
as well as an implementation of the relativistic MP2 method utilizing Cholesky-decomposed density matrices.\cite{helmich-parisRelativisticCholeskydecomposedDensity2019}

In this paper, we present a new implementation of the X2CAMF-CCSD(T) and EOM-CCSD schemes using Cholesky decomposed two-electron integrals, aiming at lifting the storage bottlenecks and extending the applicability of these methods to medium-sized molecules.
The present CD-based implementation for relativistic CC methods uses AO-based algorithms to avoid the construction of MO two-electron integrals involving three and four virtual indices.
This improves the efficiency since the number of virtual spinors in a spinor-based two-component CC calculation is almost twice the number of AOs.
The Theory Section begins with a general consideration of the implementation, and involves as well the working equations for the CD-based X2CAMF-CCSD(T) and EOM-CCSD methods.
Subsequently, in Theory sec \ref{compSec}, we provide the computational details of our benchmark and illustrative calculations. 
The computational results are discussed in Section \ref{resultSec}. 
Finally, we conclude with a brief summary and an outlook on  future developments concerning relativistic CC methods in Section \ref{summarySec}. 

\section{Theory}

\label{theorySec}

In the Theory Section, we start with a review of relativistic CC methodologies and discuss the associated computational challenge. 
Then we present some general considerations concerning the implementation of a CD-based relativistic CC program. 
We then address the major difference between the present implementation and the available CD-based implementation for non-relativistic CC methods.
This is followed by the working equations for our CD-based X2CAMF-CCSD(T) and EOM-CCSD methods.

\subsection{General consideration}

\label{subsec:General-consideration}
In CC theory, the wave function is obtained by acting with the exponential of a cluster operator consisting of excitation operators on a reference wave function
\begin{equation}
\label{cc ene}
     | \Psi_\mathrm{CC} \rangle = e^{T} | \Psi_{0} \rangle,
\end{equation}
where $| \Psi_{0} \rangle$ is usually taken as the HF determinant.
The CC energy and amplitudes are determined by projecting Eqn. (\ref{cc ene}) from the left side with the reference and excited determinants.
For example, in the CCSD method \cite{purvisiiiFullCoupledCluster1982}, the energy and amplitude equations are given by
\begin{equation}
   \langle \Psi_{0} | \bar{H} | \Psi_{0} \rangle = E_\mathrm{CC} ,
\end{equation}
and
\begin{equation}
   \langle \Psi_{i}^{a} | \bar{H} | \Psi_{0} \rangle = 0,
\end{equation}
\begin{equation}
   \langle \Psi_{ij}^{ab} | \bar{H} | \Psi_{0} \rangle = 0 ,
\end{equation}
where $| \Psi_{i}^{a} \rangle$ and $| \Psi_{ij}^{ab} \rangle$ are singly and doubly excited determinants, respectively.
We follow here and in the following the usual convention that $i,j,\cdots $ and $a,b,\cdots$ denote occupied and virtual orbitals, respectively.
The CCSD amplitude equations are well documented in the literature.\cite{stantonDirectProductDecomposition1991,gaussCoupledclusterCalculationsNuclear1995}
In a traditional CCSD calculation, the computational bottleneck usually is the ``particle-particle ladder term'', i.e., $\frac{1}{2}\sum_{ef} \langle ab || ef\rangle t_{ij}^{ef}$, while the storage bottleneck is due to the MO two-electron integrals, especially the $\langle ab||cd\rangle$ integrals.
These bottlenecks are even more severe in a spinor-based relativistic CCSD calculation, with spin-orbit coupling included in the orbitals, because of the spin-symmetry breaking.
The MO coefficients are generally complex-valued in a spinor-based calculation
and the number of molecular spinors is the sum of the numbers of $\alpha$ and $\beta$ molecular orbitals.
The storage of the MO two-electron integrals thus becomes 
an order of magnitude more expensive than a corresponding non-relativistic calculations.

Consider a calculation with 550 basis functions and the correlation of around 100 electrons ($n_{\mathrm o}=100$) and 1000 virtual spinors ($n_{\mathrm v}=1000$).
Since the number of virtual spinors is substantially larger than occupied spinors, when storing integrals in blocks,
the amount of memory or disk space required increases as more virtual indices are involved.
The storage requirements for blocks of AO and MO two-electron integrals are summarized in Table \ref{memint}.
\begin{table}
\centering %
\begin{tabular}{cc}
\hline \hline
Integral type  & Storage (in GB) \tabularnewline
\hline 
$(\mu\nu|\sigma\rho)$  & 85 \tabularnewline
$\langle ab||cd\rangle$  & 3725 \tabularnewline
$\langle ab||ci\rangle$  & 745 \tabularnewline
$\langle ai||bj\rangle$  & 149 \tabularnewline
$\langle ab||ij\rangle$  & 37.2 \tabularnewline
$\langle ai||jk\rangle$  & 7.5 \tabularnewline
$\langle ij||kl\rangle$  & 1.5 \tabularnewline
\hline \hline
\end{tabular}\caption{Storage requirements for two-electron integrals in an X2CAMF-CC calculation of an example system with 550 AO basis functions and 100 electrons.}
\label{memint} 
\end{table}
The storage of MO two-electron integrals with four virtual indices is usually not feasible with regular computational resources. 
It is therefore crucial to avoid the explicit construction and storage of the $\langle ab||cd\rangle$ integrals as well as of any intermediates of the same size. 
AO-based algorithms, in which the ladder term is evaluated by partially transforming the $t_{2}$ amplitudes to AO representation and contracting the transformed $t_{2}$ amplitudes with two-electron AO integrals 
\cite{ahlrichsDirectDeterminationPair1975,meyerTheorySelfConsistent1976,popleTheoreticalModelsIncorporating1976a,hampelComparisonEfficiencyAccuracy1992,kochDirectAtomicOrbital1994,gaussCoupledclusterCalculationsNuclear1995,scuseriaLinearScalingCoupled1999}, in combination with the exploitation of spin symmetry in the AO integrals \cite{liuTwocomponentRelativisticCoupledcluster2018,asthanaExactTwocomponentEquationofmotion2019}
significantly improve the efficiency of relativistic two-component CCSD and EOM-CCSD calculations.
The next storage bottleneck, as indicated in Table \ref{memint}, arises from storing the $\langle ab||ci\rangle$ integrals and intermediates of the same structure. 
To extend the applicability of two-component CC methods to medium-sized molecules targeted in the present work, it is necessary to avoid the storage of the $\langle ab||ci\rangle$-type integrals and intermediates as well. This motivates the present development of a CD-based implementation.

The CD of AO two-electron integrals can be written as
\begin{equation}
(\mu\nu|\sigma\rho)\approx\sum_{P}^{n_{\mathrm{CD}}}L_{\mu\nu}^{P}L_{\sigma\rho}^{P},
\end{equation}
where $L_{\mu\nu}^{P}$ denotes the Cholesky vectors and $n_{\mathrm{CD}}$ is the total number of Cholesky vectors. 
The Cholesky vectors are generated using an iterative procedure until the largest updated diagonal element $(\mu\nu|\nu\mu)$ is smaller than a predefined Cholesky threshold. 
The predefined threshold reflects the tolerance for linear dependence of Cholesky vectors in the Coulomb metric and sets an upper bound for the truncation errors.\cite{pedersenDensityFittingAuxiliary2009}
The anti-symmetrized MO two-electron integrals can be constructed from the Cholesky vectors in the MO basis $L_{pq}^{P}=\sum_{\mu\nu}C_{\mu p}^{\ast}L_{\mu\nu}^{P}C_{\nu q}$ on the fly instead of being stored on the disk, i.e.,
\begin{equation}
\langle pq||rs\rangle=\sum_{P}\left(L_{pr}^{P}L_{qs}^{P}-L_{ps}^{P}L_{qr}^{P}\right).
\end{equation}
Now we discuss the major difference in the strategy in handling the $\langle ab||cd \rangle$ integrals in the present implementation compared to available CD-based implementations for non-relativistic CCSD calculations. 
Since the number of $\alpha$ or $\beta$ virtual orbitals is smaller than that of the AOs in a non-relativistic calculation, it is more efficient to construct $\langle ab || cd \rangle$ from $L^P_{ab}$ than to construct the $\langle \mu\nu || \sigma\rho \rangle$ integrals from $L^P_{\mu\nu}$.
The non-relativistic CD-based CCSD schemes thus evaluate $\langle ab || cd \rangle$ and $\langle ab || ci\rangle$ on the fly using $L^P_{ab}$ and $L^P_{ai}$.
In contrast, in spinor-based relativistic CCSD calculations, the number of virtual spinors is almost twice the number of AOs.
The full construction/evaluation of the $\langle ab||cd\rangle$ and the $\langle ab||ci\rangle$ integrals thus is much more expensive than those in non-relativistic CC calculations. 
We hence decided to employ AO-based algorithms and to work with the on-the-fly construction of $\langle \mu\nu||\sigma\rho\rangle$ integrals. 
As shown in the working equations, the contractions involving these integrals in X2CAMF-CCSD and EOM-CCSD calculations can be rearranged in such a way that the Cholesky vectors are contracted first with other quantities, e.g., with the $t_{1}$-amplitudes, to reduce the dimension or to be reformulated into the AO-based algorithms. 
We mention that the storage and the construction of integrals and intermediates involving two or fewer virtual indices are not a primary bottleneck in the present study. 
These integrals thus are kept explicitly in the working equations.

\subsection{X2CAMF-CCSD(T) with Cholesky-decomposed integrals}

\label{gscc}

In our previous implementation of the X2CAMF-CCSD(T) method,\cite{liuTwocomponentRelativisticCoupledcluster2018} the $\langle ab||cd\rangle$ integrals are not stored and the ladder term is calculated using AO-based algorithms while the $\langle ab||ci\rangle$ integrals are stored on disk and the contractions involving these integrals are performed with an out-of-core algorithm. 
In our CD-based X2CAMF-CCSD(T) implementation, contractions involving $\langle ab||cd\rangle$- and $\langle ab||ci\rangle$-type integrals are reformulated using Cholesky vectors. 
The resulting $T_{1}$ equations can be written as
\begin{equation}
\begin{aligned}0 & =f_{ia}+\sum_{e}t_{i}^{e}\widetilde{f}_{ae}-\sum_{m}t_{m}^{a}\widetilde{f}_{mi}+\sum_{me}t_{im}^{ae}\widetilde{f}_{me}-\sum_{nf}t_{n}^{f}\langle na||if\rangle\\
 & -\sum_{Pf}L_{af}^{P}\sum_{me}t_{im}^{ef}L_{me}^{P}-\frac{1}{2}\sum_{men}t_{mn}^{ae}\langle nm||ei\rangle.
\end{aligned}
\label{t1}
\end{equation}
The replacement of the original contraction with one using Cholesky vectors introduces additional floating-point operations.
For example, the number of floating-point operations involved in the $\sum_{Pf}L_{af}^{P}\sum_{me}t_{im}^{ef}L_{me}^{P}$ term of the $T_{1}$ equations scales as $O(n_{\mathrm v}^{2}n_{\mathrm o}^{2}n_{\mathrm{CD}})$.
It is more expensive than the original contraction which only scales as $O(n_{\mathrm v}^{3}n_{\mathrm o}^{2})$. 
On the other hand, since this term is not computationally intensive, the increase in floating-point operations has negligible effects and is outweighed by the benefit of removing the explicit $\langle ab||ci\rangle$ integrals from the calculation.

As discussed in Sect. \ref{subsec:General-consideration}, the explicit construction of the $\langle ab||cd\rangle$ and $\langle ab||ci\rangle$ integrals from the Cholesky vectors in a spinor-based relativistic CCSD calculation
is significantly more expensive than the construction of the AO two-electron integrals, $\langle \mu\nu||\sigma\rho\rangle$.
We thus reformulated the corresponding contractions in the $T_2$ equations using AO-based algorithms. 
The $T_{2}$ equations now read
\begin{equation}
\begin{aligned}0 & =\langle ij||ab\rangle+P_{-}(ab)\sum_{e}t_{ij}^{ae}(\widetilde{f}_{be}-\frac{1}{2}\sum_{m}t_{m}^{b}\widetilde{f}_{me})-P_{-}(ij)\sum_{m}t_{im}^{ab}(\widetilde{f}_{mj}+\frac{1}{2}\sum_{e}t_{j}^{e}\widetilde{f}_{me})\\
 & +\frac{1}{2}\sum_{mn}\tau_{mn}^{ab}\widetilde{W}_{mnij}+\frac{1}{2}\sum_{\mu\nu}C_{\mu a}^{\ast}C_{\nu b}^{\ast}\widetilde{\widetilde{\tau}}_{ij}^{\mu\nu}+P_{-}(ij)P_{-}(ab)\sum_{me}(t_{im}^{ae}\widetilde{W}_{mbej}-t_{i}^{e}t_{m}^{a}\langle mb||ej\rangle)\\
 & +P_{-}(ij)P_{-}(ab)\sum_{P}L_{bj}^{P}\sum_{e}t_{i}^{e}L_{ae}^{P}-P_{-}(ab)\sum_{m}t_{m}^{a}(\langle mb||ij\rangle+\frac{1}{2}\sum_{\mu\nu}C_{\mu m}^{\ast}C_{\nu b}^{\ast}\widetilde{\widetilde{\tau}}_{ij}^{\mu\nu}),
\end{aligned}
\label{t2}
\end{equation}
where $C_{\mu p}$ denotes the MO coefficients and 
\begin{equation}
\widetilde{\widetilde{\tau}}_{ij}^{\mu\nu}=\sum_{\sigma\rho}\langle\mu\nu||\sigma\rho\rangle\sum_{ef}C_{\sigma e}C_{\rho f}\tau_{ij}^{ef}
\end{equation}
are the intermediates in the AO-based algorithms. $P_{-}(pq)=1-\widetilde{p}(pq)$ is the anti-symmetrization operator, in which $\widetilde{p}(pq)$ interchanges the labels $p$ and $q$. 
The spin-factorized formulae of the $\widetilde{\widetilde{\tau}}_{ij}^{\mu\nu}$ are given as
\begin{equation}
\begin{aligned}
    \widetilde{\widetilde{\tau}}_{ij}^{\mu^{s}\nu^{s}}
    &=\sum_{\sigma^{s}\rho^{s}}\langle\mu^{s}\nu^{s}||\sigma^{s}\rho^{s}\rangle\sum_{ef}C_{\sigma^{s}e}C_{\rho^{s}f}\tau_{ij}^{ef} \\
    &=\sum_{\sigma^{s}\rho^{s}}
    \sum_{P}^{n_{\mathrm{CD}}}\left(L_{\mu^{s}\sigma^{s}}^{P}L_{\nu^{s}\rho^{s}}^{P} - L_{\mu^{s}\rho^{s}}^{P}L_{\nu^{s}\sigma^{s}}^{P}\right)
    \sum_{ef}C_{\sigma^{s}e}C_{\rho^{s}f}\tau_{ij}^{ef},\text{\ }s=\alpha\text{ or }\beta
\end{aligned}
\end{equation}
and
\begin{equation}
\begin{aligned}
    \widetilde{\widetilde{\tau}}_{ij}^{\mu^{\alpha}\nu^{\beta}}
    &=\sum_{\sigma^{\alpha}\rho^{\beta}}  (\mu\sigma|\nu\rho) \sum_{ef}C_{\sigma^{\alpha}e}C_{\rho^{\beta}f}\tau_{ij}^{ef}\\
    &=\sum_{\sigma^{\alpha}\rho^{\beta}}
    \left(\sum_{P}^{n_{\mathrm{CD}}}L_{\mu^{\alpha}\sigma^{\alpha}}^{P}L_{\nu^{\beta}\rho^{\beta}}^{P}\right)
    \sum_{ef}C_{\sigma^{\alpha}e}C_{\rho^{\beta}f}\tau_{ij}^{ef}.
\end{aligned}
\end{equation}

The intermediates $\widetilde{f}_{mi}$, $\widetilde{f}_{me}$, $\widetilde{W}_{mnij}$, $\widetilde{\tau}_{ij}^{ab}$, and $\tau_{ij}^{ab}$ have the same definition as those in the standard non-relativistic CCSD equations\cite{gaussCoupledclusterCalculationsNuclear1995} and the corresponding X2CAMF-CCSD equations\cite{liuTwocomponentRelativisticCoupledcluster2018}
while the intermediates $\widetilde{f}_{ae}$ and $\widetilde{W}_{mbej}$ are slightly modified for the CD-based implementations. 
For the sake of completeness, we provide a summary of the expressions for these intermediates in Table~\ref{ftW}.
The present working equations differ from those for CD-based non-relativistic CCSD given in Ref. ~\onlinecite{epifanovskyGeneralImplementationResolutionoftheidentity2013} in that we make explicit use of MO two-electron integrals with two or fewer virtual indices and use AO-based algorithms to avoid the construction of the $\langle ab||cd\rangle$ and $\langle ab||ci\rangle$ integrals.
The latter is essential for the efficiency of CD-based relativistic two-component CCSD calculations.
This also means that $\widetilde{W}_{mbej}$, about four times the size of $T_2$, is the largest intermediate that appears in the present implementation.

\begin{table}
\centering{}\caption{Definitions of the intermediates in the CD-based X2CAMF-CCSD 
Eqns. (\ref{t1}) and (\ref{t2}).}
\label{ftW} %
\begin{tabular}{l}
\hline \hline
Intermediates in the CD-based X2CAMF-CCSD equations \tabularnewline
\hline 
$\widetilde{f}_{mi}=f_{mi}-\frac{1}{2}\sum_{e}f_{me}t_{i}^{e}+\sum_{en}t_{n}^{e}\langle mn||ie\rangle+\frac{1}{2}\sum_{nef}\widetilde{\tau}_{in}^{ef}\langle mn||ef\rangle$ \tabularnewline
$\widetilde{f}_{ae}$ = $f_{ae}-\frac{1}{2}\sum_{m}f_{me}t_{m}^{a}-\frac{1}{2}\sum_{mnf}\widetilde{\tau}_{mn}^{af}\langle mn||ef\rangle$ \tabularnewline
~~~~~$+\left(\sum_{P}L_{ae}^{P}\sum_{mf}t_{m}^{f}L_{mf}^{P}-\sum_{Pm}L_{me}^{P}\sum_{f}t_{m}^{f}L_{af}^{P}\right)$ \tabularnewline
$\widetilde{f}_{me}$ = $f_{me}+\sum_{nf}t_{n}^{f}\langle mn||ef\rangle$ \tabularnewline
$\widetilde{W}_{mnij}$ = $\langle mn||ij\rangle+P_{-}(ij)\sum_{e}t_{j}^{e}\langle mn||ie\rangle+\frac{1}{2}\tau_{ij}^{ef}\langle mn||ef\rangle$ \tabularnewline
$\widetilde{W}_{mbej}$ = $\langle mb||ej\rangle-\sum_{n}t_{n}^{b}\langle mn||ej\rangle-\sum_{nf}(\frac{1}{2}t_{jn}^{fb}+t_{j}^{f}t_{n}^{b})\langle mn||ef\rangle$\tabularnewline
~~~~~~~~~$+\sum_{P}\left(L_{me}^{P}\sum_{f}t_{j}^{f}L_{bf}^{P}-L_{be}^{P}\sum_{f}t_{j}^{f}L_{mf}^{P}\right)$\tabularnewline
$\widetilde{\tau}_{ij}^{ab}$ = $t_{ij}^{ab}+\frac{1}{2}(t_{i}^{a}t_{j}^{b}-t_{i}^{b}t_{j}^{a})$ \tabularnewline
$\tau_{ij}^{ab}$ = $t_{ij}^{ab}+t_{i}^{a}t_{j}^{b}-t_{i}^{b}t_{j}^{a}$ \tabularnewline
\hline \hline
\end{tabular}
\end{table}

With the converged amplitudes $t_{i}^{a}$ and $t_{ij}^{ab}$, one can calculate the CCSD energy \cite{purvisiiiFullCoupledCluster1982} and the (T) correction\cite{raghavachariFifthorderPerturbationComparison1989} as 
\begin{equation}
E_{\text{CCSD}}=\sum_{ia}t_{i}^{a}f_{ia}+\frac{1}{4}\sum_{ijab}\tau_{ij}^{ab}\langle ij||ab\rangle
\end{equation}
and 
\begin{equation}
E_{\text{(T)}}=\frac{1}{36}\sum_{ijkabc}t(c)_{ijk}^{abc}D_{ijk}^{abc}(t(c)_{ijk}^{abc}+t(d)_{ijk}^{abc}),
\end{equation}
where $t(d)_{ijk}^{abc}=D_{ijk}^{abc}P(i/jk)P(a/bc)t_{i}^{a}\langle bc||jk\rangle$,
$D_{ijk}^{abc}=1/(f_{ii}+f_{jj}+f_{kk}-f_{aa}-f_{bb}-f_{cc})$, and
\begin{equation}
t(c)_{ijk}^{abc}=D_{ijk}^{abc}(\sum_{e}P(i/jk)P(a/bc)t_{jk}^{ae}\sum_{P}\left(L_{be}^{P}L_{ci}^{P}-L_{bi}^{P}L_{ce}^{P}\right)+\sum_{m}P(i/jk)P(a/bc)t_{mi}^{bc}\langle jk||ma\rangle).
\end{equation}
$P(p/qr)$ are given as $1-\widetilde{p}(pq)-\widetilde{p}(pr)$.
The (T) correction given above introduces additional floating-point operations
scaling as $O(n_{\mathrm o}^{3}n_{\mathrm v}^{3}n_{\mathrm{CD}})$ and thus is more expensive
than a standard (T) calculation, with the computational time around 2-4 times longer in practice. 
It would be of interest to explore other decomposition schemes 
\cite{cacheiroCCSDModelCholesky2011,nagyOptimizationLinearscalingLocal2017} and local approximations \cite{schutzLocalPerturbativeTriples2000,guoCommunicationImprovedLinear2018,guoLinearScalingPerturbative2020} to accelerate the calculation of the (T) corrections in future work. 

\subsection{X2CAMF-EOM-CCSD with Cholesky-decomposed integrals}

\label{eom} In our previous X2CAMF-EOM-CCSD implementation, \cite{asthanaExactTwocomponentEquationofmotion2019} all blocks of the similarity-transformed Hamiltonian ($\bar{H}$), except $\bar{H}_{ab,cd}$, are pre-calculated after the ground-state CCSD calculation and stored on disk. 
In the present CD-based implementation, the need of storing $\bar{H}_{ai,bc}$ and $\bar{H}_{ab,ci}$ is eliminated by calculating these terms on the fly in each iteration using the Cholesky vectors. 
Similar to the ground-state CCSD calculation, the contractions involving the $\langle ab||cd\rangle$ and $\langle ab||ci\rangle$ integrals can be reformulated using AO-based algorithms.
The residue vectors for the CD-based X2CAMF-EOM-CCSD secular equations are given by
\begin{equation}
\begin{aligned}\widetilde{r}_{i}^{a}= & -E_{\text{ex}}r_{i}^{a}+\sum_{c}\bar{H}_{ac}r_{i}^{c}-\sum_{k}\bar{H}_{ki}r_{k}^{a}+\sum_{kc}\bar{H}_{kc}r_{ik}^{ac}\\
+ & \sum_{kc}\bar{H}_{akic}r_{k}^{c}+\sum_{Pc}L_{ac}^{P}\sum_{kd}L_{kd}^{P}r_{ik}^{cd}-\frac{1}{2}\sum_{kcd}r_{ik}^{cd}\sum_{l}t_{l}^{a}\langle lk||cd\rangle-\frac{1}{2}\sum_{kld}\bar{H}_{klid}r_{kl}^{ad}
\end{aligned}
\end{equation}
and 
\begin{equation}
\begin{aligned}\widetilde{r}_{ij}^{ab}= & -E_{\text{ex}}r_{ij}^{ab}+P_{-}(ab)\sum_{e}\bar{H}_{be}r_{ij}^{ae}-P_{-}(ij)\sum_{k}\bar{H}_{kj}r_{ik}^{ab}\\
+ & \frac{1}{2}\sum_{kl}\bar{H}_{klij}r_{kl}^{ab}+P_{-}(ij)P_{-}(ab)\sum_{kc}\bar{H}_{bkjc}r_{ik}^{ac}\\
+ & \frac{1}{2}\sum_{\mu\nu}C_{\mu a}^{\ast}C_{\nu b}^{\ast}\widetilde{\widetilde{r}}_{ij}^{\mu\nu}-\frac{1}{2}P_{-}(ab)\sum_{m}t_{m}^{b}\sum_{\mu\nu}C_{\mu a}^{\ast}C_{\nu m}^{\ast}\widetilde{\widetilde{r}}_{ij}^{\mu\nu}+\frac{1}{4}\sum_{mn}\tau_{mn}^{ab}\sum_{ef}r_{ij}^{ef}\langle mn||ef\rangle\\
- & P_{-}(ij)\sum_{m}t_{mj}^{ab}\sum_{c}\widetilde{f}_{mc}r_{i}^{c}+\frac{1}{2}P_{-}(ij)\sum_{mn}\tau_{mn}^{ab}\sum_{c}\bar{W}_{mncj}r_{i}^{c}+P_{-}(ij)P_{-}(ab)\sum_{P}L_{bj}^{P}M_{ai}^{P}\\
+ & P_{-}(ij)P_{-}(ab)\sum_{P}M_{ai}^{P}\sum_{f}L_{bf}^{P}t_{j}^{f}-P_{-}(ij)P_{-}(ab)\sum_{m}t_{m}^{a}\sum_{c}\bar{W}_{mbcj}r_{i}^{c}\\
- & P_{-}(ij)P_{-}(ab)\left(\sum_{mf}t_{mj}^{af}\sum_{P}L_{bf}^{P}\sum_{c}L_{mc}^{P}r_{i}^{c}-\sum_{P}M_{bi}^{P}\sum_{mf}L_{mf}^{P}t_{mj}^{af}\right)\\
+ & P_{-}(ab)\sum_{d}t_{ij}^{ad}\left(-\sum_{n}t_{n}^{b}\sum_{kc}\langle nk||dc\rangle r_{k}^{c}+\sum_{P}L_{bd}^{P}\sum_{kc}L_{kc}^{P}r_{k}^{c}-\sum_{kP}L_{kd}^{P}M_{bk}^{P}\right)\\
+ & P_{-}(ab)\sum_{k}\bar{H}_{akij}r_{k}^{b}-P_{-}(ij)\sum_{l}t_{il}^{ab}\sum_{kc}\bar{H}_{klcj}r_{k}^{c}\\
+ & P_{-}(ij)\frac{1}{2}\sum_{l}t_{il}^{ab}\sum_{cdk}\langle kl||dc\rangle r_{jk}^{cd}-P_{-}(ab)\frac{1}{2}\sum_{e}t_{ij}^{ae}\sum_{kld}\langle kl||ed\rangle r_{kl}^{bd},
\end{aligned}
\end{equation}
where $E_{\text{ex}}$ is the excitation energy, $M_{ai}^{P}=\sum_{c}L_{ac}^{P}r_{i}^{c}$, and 
\begin{equation}
\widetilde{\widetilde{r}}_{ij}^{\mu\nu}=\sum_{\sigma\rho}\langle\mu\nu||\sigma\rho\rangle\sum_{ef}C_{\sigma e}C_{\rho f}r_{ij}^{ef}.
\end{equation}
The expressions for the similarity-transformed Hamiltonian
$\bar{H}$ as well as some other intermediates are summarized
in Table \ref{hbar}. 
We mention that similar AO-based algorithms can be used for the CD-based implementation of the EOM-CCSD left eigenvalue equations and the $\Lambda$-equations in CC gradient theory. 

\begin{table}
\centering{}\caption{{Definitions of the blocks of the similarity-transformed Hamiltonian $\bar{H}$
and the $\bar{W}$ intermediates.}}
\label{hbar} %
\begin{tabular}{l}
\hline \hline
$\bar{H}$ and $\bar{W}$ intermediates \tabularnewline
\hline 
$\bar{H}_{me}=\widetilde{f}_{me}$ \tabularnewline
$\bar{H}_{mi}=\widetilde{f}_{mi}+\frac{1}{2}\sum_{e}t_{i}^{e}\widetilde{f}_{me}$ \tabularnewline
$\bar{H}_{ae}=\widetilde{f}_{ae}+\frac{1}{2}\sum_{m}t_{m}^{a}\widetilde{f}_{me}$ \tabularnewline
$\bar{H}_{mnij}=\widetilde{W}_{mnij}$\tabularnewline
$\bar{H}_{mbej}=\widetilde{W}_{mbej}-\frac{1}{2}\sum_{fn}t_{jn}^{fb}\langle mn||ef\rangle$\tabularnewline
$\bar{H}_{mnie}=\langle mn||ie\rangle+\sum_{f}t_{i}^{f}\langle mn||fe\rangle$ \tabularnewline
$\bar{H}_{mbij}=\langle mb||ij\rangle-\sum_{e}\bar{H}_{me}t_{ij}^{be}-\sum_{n}t_{n}^{b}\bar{H}_{mnij}$ \tabularnewline
~~~~~~~~$+\sum_{efP}\left(L_{me}^{P}L_{bf}^{P}\right)\tau_{ij}^{ef}+P_{-}(ij)\sum_{ne}\langle mn||ie\rangle t_{jn}^{be}$\tabularnewline
~~~~~~~~$+P_{-}(ij)\sum_{e}\left(\langle mb||ej\rangle-\sum_{nf}t_{nj}^{bf}\langle mn||ef\rangle\right)$\tabularnewline
$\bar{W}_{mn,ci}=\langle mn||ci\rangle+\sum_{f}t_{i}^{f}\langle mn||cf\rangle$\tabularnewline
$\bar{W}_{mb,ci}=\langle mb||ci\rangle-\sum_{nf}t_{ni}^{bf}\langle mn||cf\rangle$\tabularnewline
~~~~~~~~$+\sum_{P}\left(L_{mc}^{P}M_{bi}^{P}-L_{bc}^{P}\sum_{f}t_{i}^{f}L_{mf}^{P}\right)$\tabularnewline
\hline \hline
\end{tabular}
\end{table}

\section{Computational details}

\label{compSec} The CD-based X2CAMF-CCSD(T) and EOM-CCSD algorithms described
in Sec. \ref{theorySec} have been implemented in the CFOUR program package.\cite{stantonCFOURCoupledclusterTechniques,matthewsCoupledclusterTechniquesComputational2020}
As illustrative examples, we have computed the ionization energies (IE) of the uranium
atom (U), the uranium-monoxide molecule (UO), and the uranium-dioxide
molecule (UO$_{2}$), and for the bond-dissociation energies ($D_{e}$)
of UO and UO$_{2}$, using the uncontracted cc-pVTZ basis set \cite{ccpvxzBtoNe} for oxygen and the uncontracted ANOTZ basis set obtained by augmenting the set of uncontracted s-, p-, d-, and f-type functions in the ANO-RCC set \cite{roosNewRelativisticANO2005a,faegrijrRelativisticGaussianBasis2001}
with the correlating functions from the cc-pVTZ-X2C set.\cite{petersonCorrelationConsistentBasis2015}
For the calculation of the vertical excitation energies (VEE) of UO$_{2}$
and the uranium dinitride molecule (UN$_{2}$), we have used contracted basis sets for the X2CAMF scheme for uranium.
The $j$-adapted basis sets used here employ the primitive Gaussian functions in the uncontracted Dyall's correlation-consistent valence triple- and quadruple-zeta basis sets (dy-pVXZ, X=T, Q) and those with core-correlating functions (dy-pCVXZ, X=T, Q) \cite{dyall5f}. 
The contraction coefficients are determined by atomic X2CAMF calculations using spherically averaged occupation numbers. 
This means that separate contraction coefficients are used for basis functions with the same orbital angular momenta but different $j$ values.
We refer to these basis sets as dy-p(C)VXZ-SO (X=T,Q).
The corresponding contraction coefficients for the spin-free relativistic calculations using spin-free X2C in its one-electron variant (SFX2C-1e) \cite{dyallInterfacingRelativisticNonrelativistic2001} have also been obtained and will be denoted as dy-p(C)VXZ-SF (X=T, Q) sets. 
The basis sets used in this paper are given in the supplementary material.
Details about the construction of these basis sets are beyond the scope of this paper and will be reported elsewhere. 
In the calculations of excitation energies of UO$_{2}$ and UN$_{2}$,
cc-pVTZ basis sets recontracted for the SFX2C-1e scheme have been used for oxygen and
nitrogen while the dy-pVTZ-SO set has been used for uranium.
The uranium 6s, 6p, 7s, 6d, 5f electrons and the 2s and 2p electrons of oxygen and nitrogen have been correlated in these calculations. 
The calculations using uncontracted basis sets have kept virtual spinors higher than 100 Hartree frozen in the CC calculations. 

For the calculations of the bond-dissociation energy of uranium hexafluoride
(\ce{UF6}), dy-pVXZ-SO and dy-pCVXZ-SO basis sets have been used
for uranium and the cc-pVXZ basis sets recontracted for the SFX2C-1e scheme
have been used for fluorine. We refer to the combination of uranium dy-pVXZ-SO set
and the fluorine cc-pVXZ set as the VXZ set and the combination of uranium dy-pCVXZ-SO set
and fluorine cc-pVXZ set as the cVXZ set. The uranium 6s, 6p, 7s, 6d,
5f electrons and the fluorine 2s, 2p electrons have been correlated
in the calculations using the VXZ basis sets. The uranium 5s, 5p,
and 5d electrons have been further correlated in the calculations
using the cVXZ basis sets. The basis-set-limit values for the CD-based
X2CAMF-CCSD results have been estimated using the cVTZ and cVQZ correlation
energies and a two-point formula \cite{basissetExtrapolation} augmented
with the X2CAMF-HF/cVQZ energy. The (T) correction {[}$\Delta$(T){]}
has been obtained using the cVTZ basis sets. 
These calculations have been performed using geometries for UF$_{5}$ ($C_{4v}$) and UF$_{6}$ ($O_{h}$) optimized at the spin-free CCSD(T)/cc-pVQZ-DK3 level reported in Ref. ~\onlinecite{petersonCorrelationConsistentBasis2015}.
The zero-point-energy corrections ($\Delta$ZPE) have been obtained by using harmonic frequencies computed at the SFX2C-1e-CCSD level using the dy-pVTZ-SF basis for uranium and the SFX2C-1e recontracted cc-pVDZ basis set for fluorine. 

For the calculations of the lowest excitation energy of the solvated uranyl ion, the dy-pVTZ-SO basis set has been used for uranium,
the SFX2C-1e recontracted cc-pVTZ basis set has been used for the oxygen atoms in the uranyl ion, 
and the SFX2C-1e recontracted cc-pVDZ basis sets have been used for the oxygen and hydrogen atoms in the water molecules forming the solvation shells. 
The uranium 6s, 6p, 7s, 6d, 5f electrons, oxygen 2s and 2p electrons, and hydrogen 1s electron have been correlated in these calculations.
For the scalar-relativistic SFX2C-1e calculations, the dy-pVTZ-SO basis set for uranium was replaced by the corresponding dy-pVTZ-SF set. 
The molecular structures of \ce{UO2^2+}, \ce{UO2^2+(H2O)4} ($D_{4h}$), \ce{UO2^2+(H2O)5} ($D_{5h}$), and \ce{UO2^2+(H2O)12} ($D_{4h}$p) are taken as those in Ref. ~\onlinecite{sibouletTheoreticalStudyUranyl2006}.
The Gaussian nuclear model \cite{visscherDiracfockAtomicElectronic1997} has been used for all the calculations.
The present calculations have not used double-group symmetry to accelerate the calculations. 

\section{Results and discussion}

\label{resultSec} 

\subsection{Benchmark calculations for uranium-containing molecules}

\begin{figure}
\begin{subfigure}{0.5\textwidth} \includegraphics[width=1\linewidth]{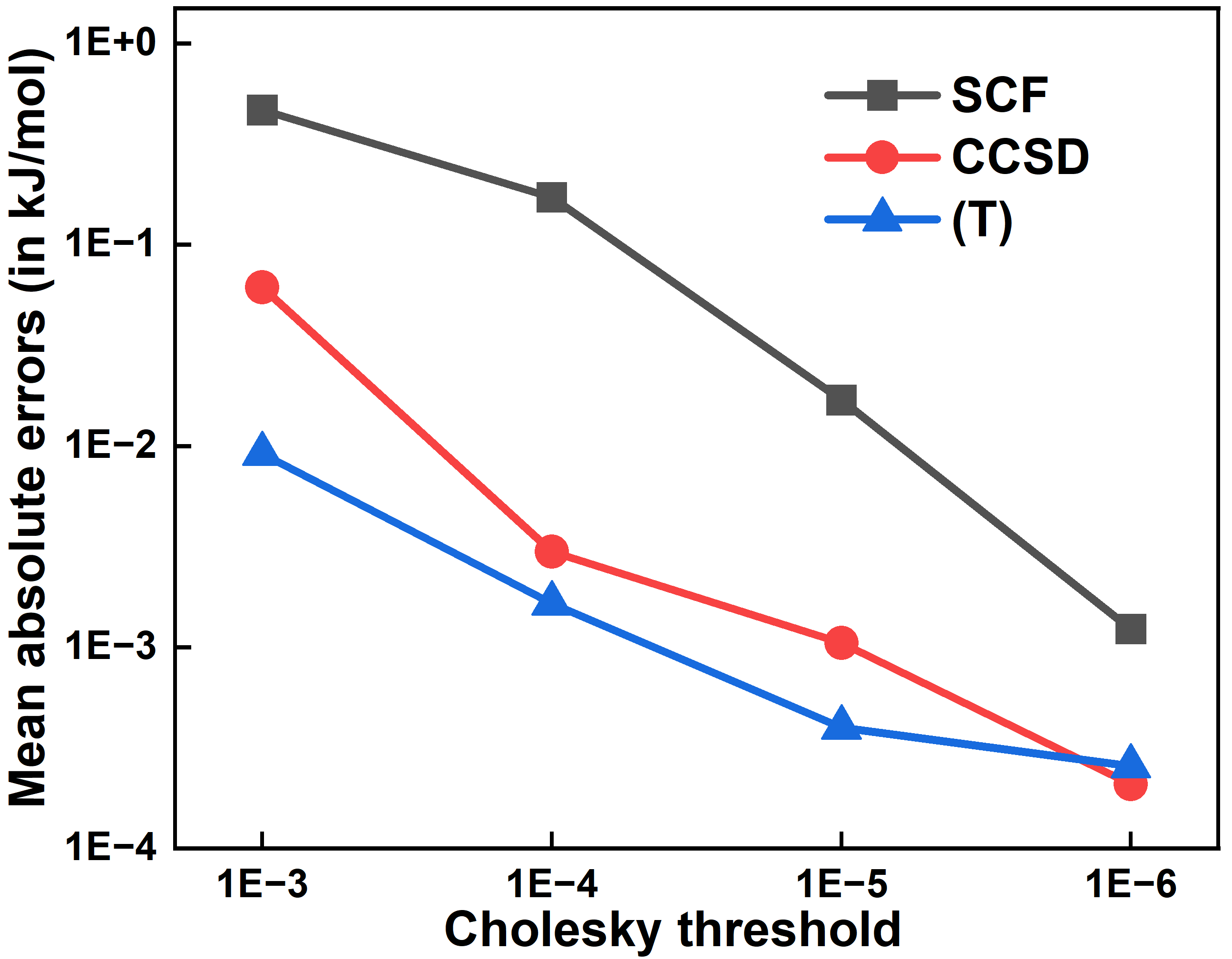}
\caption{Mean absolute errors for IE and $D_{e}$.}
\label{bchmkg} 
\end{subfigure}\hspace*{\fill} 
\begin{subfigure}{0.5\textwidth} \includegraphics[width=1\linewidth]{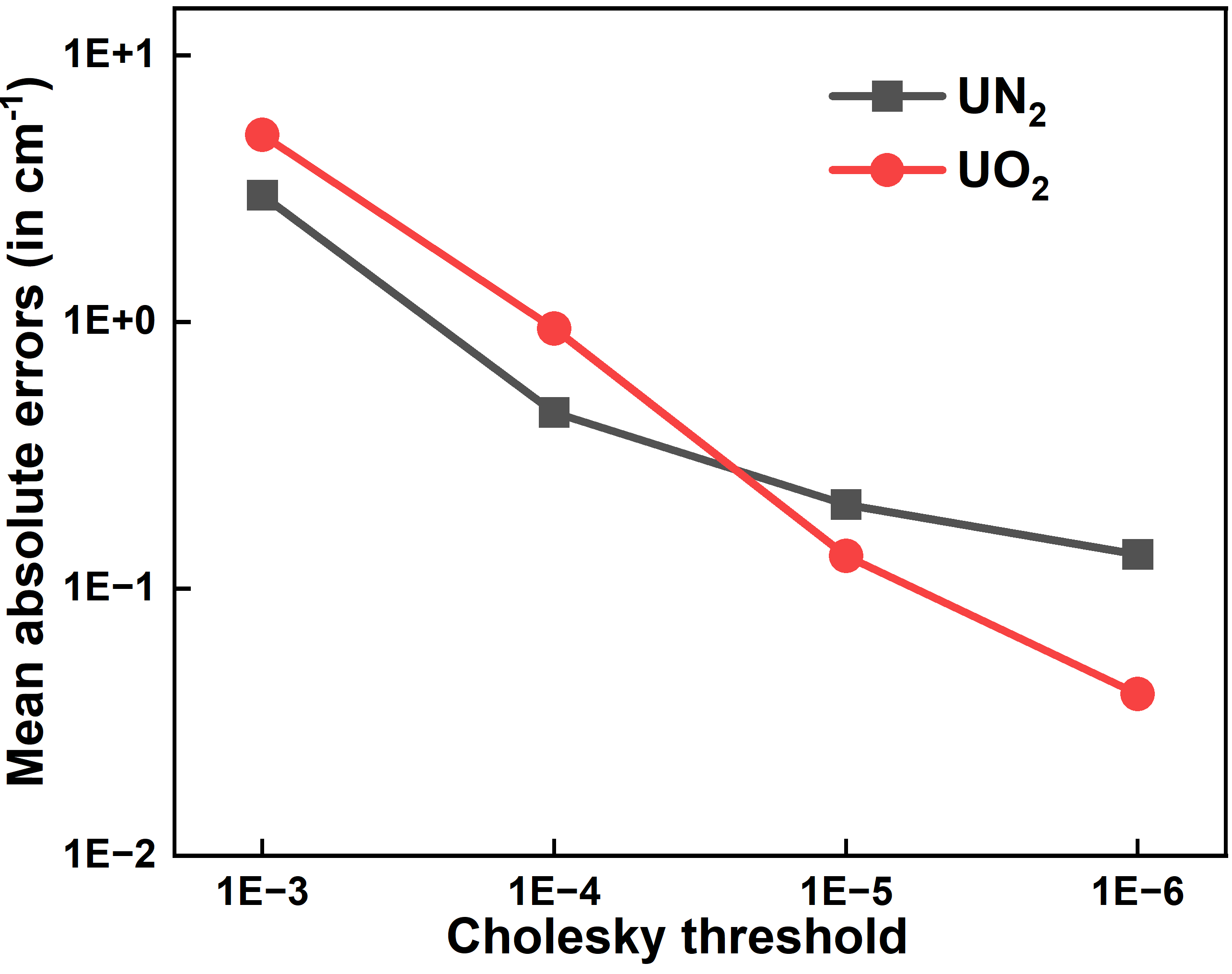}
\caption{Mean absolute errors for VEE.}
\label{bchmke} 
\end{subfigure}
\\
\begin{subfigure}{0.5\textwidth} \includegraphics[width=1\linewidth]{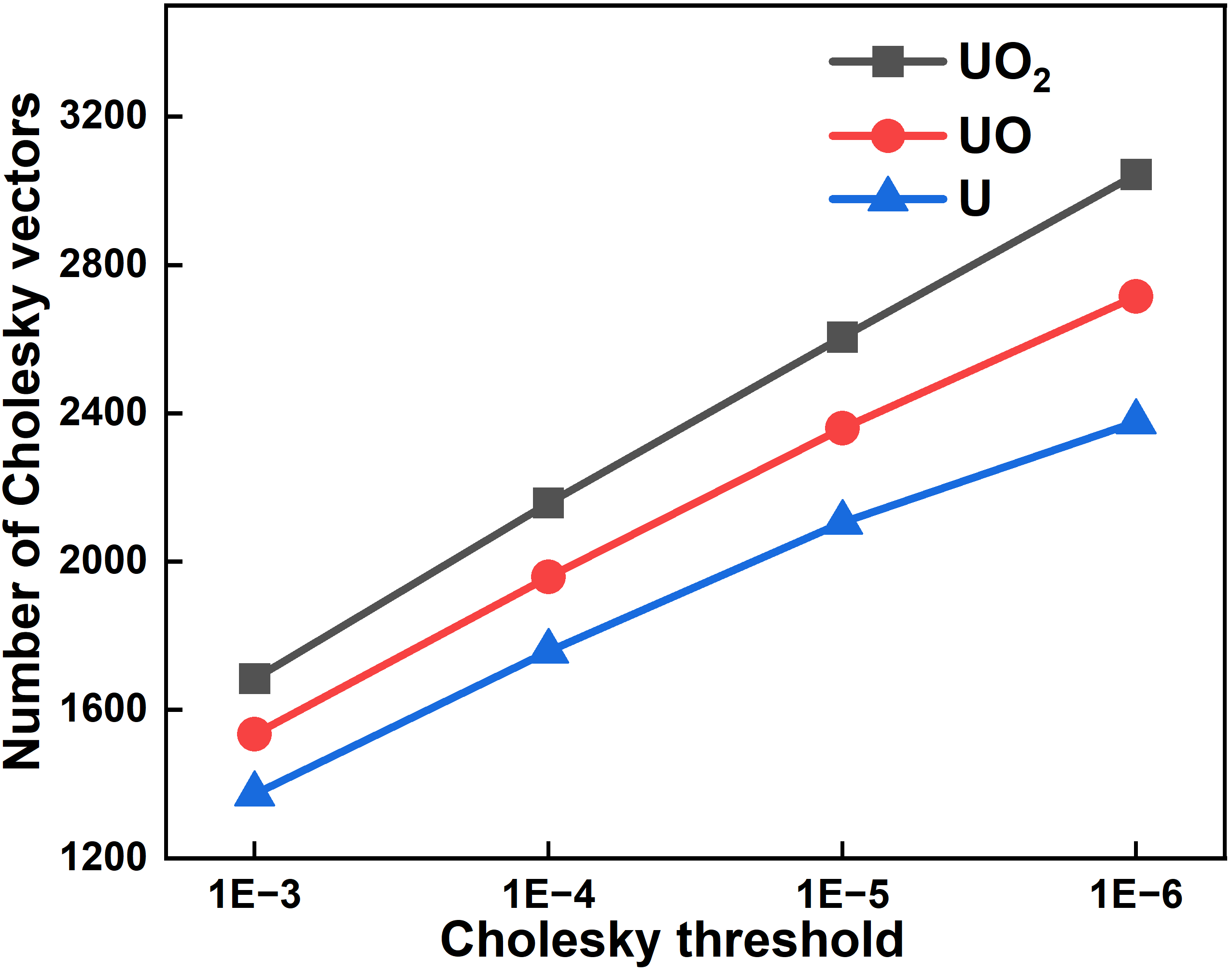}
\caption{Number of Cholesky vectors.}
\label{bchmkn} 
\end{subfigure}
\caption{Mean absolute errors for the ionization energies (IE), bond-dissociation energies ($D_{e}$), lowest-lying eighteen vertical excitation energies (VEE), and number of Cholesky vectors for uranium-containing small molecules with respect to
Cholesky decomposition thresholds of $10^{-\delta}$ ($\delta=3,4,5,6$).}
\label{bchmk} 
\end{figure}

To benchmark the accuracy of the CD-based X2CAMF-CCSD(T) and EOM-CCSD calculations, ionization energies (IE) of U, UO, and UO$_{2}$, bond-dissociation energies ($D_{e}$) of UO and UO$_{2}$,
and the first eighteen vertical excitation energies (VEE) of UO$_{2}$
and UN$_{2}$ have been calculated with Cholesky thresholds of $10^{-\delta}$
($\delta=3, 4, 5, 6$) and compared with the results obtained in corresponding traditional calculations without approximating the two-electron integrals.
The reference values for IEs and $D_{e}$'s have been taken from Ref. ~\onlinecite{zhangRouteChemicalAccuracy2022}.
The convergence patterns of the computed IEs, $D_{e}$'s, and VEEs, as well as the number of Cholesky vectors as a function of the Cholesky threshold are plotted in Fig. \ref{bchmk}.

The errors introduced by Cholesky decomposition are generally insignificant. 
The total mean absolute errors for the IE and $D_{e}$ are 0.43, 0.17, 0.02, and less than 0.01 kJ/mol for Cholesky thresholds of $10^{-\delta}$ with $\delta=3, 4, 5, 6$, respectively. 
Among the benchmark set, the maximum absolute error is observed to be 1.1 kJ/mol for $D_{e}$(UO) with a $10^{-3}$ Cholesky threshold. 
By further decreasing the Cholesky threshold to $10^{-5}$, the maximum absolute error reduces to 0.04 kJ/mol, which is adequate for computational thermochemistry aiming at sub-chemical accuracy. 
We note that both the storage requirements and floating-point operations in the X2CAMF-CC calculations scale linearly with the number of Cholesky vectors. 
Since the number of Cholesky vectors increases logarithmically with respect to the Cholesky threshold as shown in Fig. \ref{bchmkn}, one can systematically improve the computational accuracy with only a moderate increase of computational cost.

The CD errors for excitation energies are expected to be less significant because of error cancellation,
since the CD approximations treat ground state and excited states on an equal footing.
The mean absolute errors for excitation energies introduced by a $10^{-3}$ Cholesky threshold are less than 5 cm$^{-1}$. The maximum error of 12 cm$^{-1}$ was observed for an excited state of \ce{UO2} with an excitation energy of 2.2 eV. 
When decreasing the Cholesky threshold to $10^{-4}$, the maximum error is reduced to 2 cm$^{-1}$, becoming negligible for practical applications. 
Although errors from Cholesky decomposition may increase with the size of systems, a Cholesky threshold of $10^{-4}$ is a balanced choice for calculations of both ground state and excited states.

It is worth mentioning that the errors introduced by CD mainly come from the mean-field step, as shown in Fig. \ref{bchmkg}. 
One can, in principle, reduce the errors by using a hybrid scheme, in which a standard Hartree-Fock calculation without CD is performed to generate the MOs, followed by an electron-correlation calculation with CD integrals. 
However, although the hybrid scheme significantly reduces the total errors, it does not improve the accuracy of correlation energies and may lead to more complications when evaluating energy derivatives. 
Therefore, we will not use the hybrid scheme, especially considering that the normal CD scheme is already very accurate.

\subsection{Bond-dissociation energy of uranium hexafluoride}

To demonstrate the capability of CD-based X2CAMF-CC methods, we calculated the bond-dissociation energy of uranium hexafluoride (\ce{UF6}) via the reaction energy of \ce{UF6 -> UF5 + F}.  
A Cholesky threshold of $10^{-5}$ has been used for these calculations.
The calculated bond-dissociation energy with a detailed account of individual contributions are summarized in Table \ref{UFx}. 
The computational resources required for these CD-based X2CAMF-CCSD calculations are summarized in Table \ref{UF cost}.

\begin{table}
\caption{Calculated bond-dissociation energy (in kJ/mol) of \ce{UF6}}
\begin{tabular}{cc}
\hline \hline
method & bond dissociation energy\tabularnewline
\hline 
X2CAMF-CCSD/cVTZ & 249.8\tabularnewline
X2CAMF-CCSD/cVQZ & 245.7\tabularnewline
X2CAMF-CCSD/cV$\infty$Z & 240.9\tabularnewline
$\Delta$(T) & 61.5\tabularnewline
$\Delta$ZPE & -7.8\tabularnewline
Total & 294.6\tabularnewline
\hline
\multirow{2}{*}{Experimental} & 313$\pm$15$^{a}$\tabularnewline
 & 294$\pm$8\cite{hildenbrandRedeterminationThermochemistryGaseous1991}\tabularnewline
\hline \hline
\label{UFx}
\end{tabular}\\
$a$. Calculated from the enthalpy of formation for \ce{UF4}, \ce{UF5},
\ce{UF6} \cite{oecdSecondUpdateChemical2021}, and F \cite{ruscicIntroductionActiveThermochemical2004}
\end{table}

Similar to the case of uranium-containing small molecules, \cite{brossMultireferenceConfigurationInteraction2015,zhangRouteChemicalAccuracy2022} it is essential to account for the correlation of semi-core electrons, i.e., the uranium 5s, 5p, and 5d electrons, when aiming at an accurate estimate for the basis-set-limit value of the electron-correlation contribution to the dissociation energy of \ce{UF6}.
The electron-correlation contributions with semi-core electrons correlated are 253.5, 246.9, and 242.1 kJ/mol for the cVTZ, cVQZ, and cV$\infty$Z basis sets, respectively. 
In contrast, the corresponding contributions are 254.2, 254.4, and 254.5 kJ/mol with the VTZ, VQZ, and V$\infty$Z basis sets. 
The corrections from basis-set effects exhibit opposite directions in cVXZ and VXZ calculations. 
The VXZ calculations without the correlation of semi-core electrons have an error of more than 10 kJ/mol in the basis-set limit.
The (T) correction contributes more than 60 kJ/mol. 
This may indicate non-negligible remaining errors from high-level correlation contributions. 

\begin{table}
\caption{Memory requirement, minimal storage requirement of two-electron atomic orbital (2e-AO) integrals with 4-fold symmetry, and the wall time with 48 processor units on Intel Xeon Gold Cascade Lake 6248R for \ce{UF5} and \ce{UF6} using our implementation of CD-based X2CAMF-CCSD methods in CFOUR.}
\begin{tabular}{ccccc}
\hline \hline
 & \multicolumn{2}{c}{\ce{UF5}}  & \multicolumn{2}{c}{\ce{UF6}} \tabularnewline
\hline
basis set & cVTZ & cVQZ & cVTZ & cVQZ \tabularnewline
Number of atomic orbitals     & 406  & 631  & 436  & 686  \tabularnewline
Number of correlated electrons& 67   & 67   & 74   & 74   \tabularnewline
Number of virtual spinors     & 571  & 1021 & 622  & 1122 \tabularnewline
Number of Cholesky vectors    & 2203 & 3460 & 2452 & 3796 \tabularnewline
Size of $\widetilde{W}_{mbej}$ (in GB) & 22 & 70 & 32 & 103  \tabularnewline
Wall time (in hour/iteration) & 0.3 & 1.3 & 0.4 & 2.1     \tabularnewline
\hline \hline
\end{tabular}
\label{UF cost}
\end{table}

The calculations using the cVTZ basis sets in Table \ref{UF cost} show that the CD-based X2CAMF-CCSD calculations correlating more than 50 electrons and 500 virtual spinors can routinely be performed on a regular computational node and can be completed within a few to 10 hours.
The calculation of \ce{UF6} using the cVQZ basis correlates 74 electrons and involves more than 1100 virtual spinors. 
It took around 2 hours for a CCSD iteration using 48 threads. 
Here holding a $\widetilde{W}_{mbej}$-type intermediate and a t2-type intermediate in fast memory requires around 100 GB RAM and 25 GB RAM, respectively; such calculations thus require a computer node with a large memory. 

\subsection{Excitation energies of aqueous uranyl ion}

\begin{figure}
\includegraphics[width=1.0\textwidth]{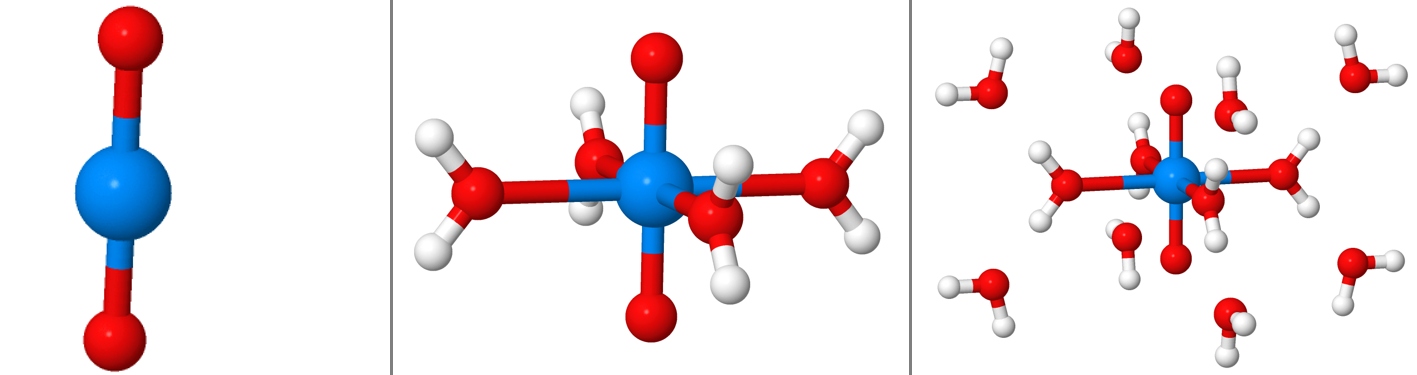}
\caption{Molecular structure for bare uranyl ion \ce{UO2^2+}, uranyl ion with the first shell \ce{UO2^2+(H2O)4}, and uranyl ion with the first two solvation shells \ce{UO2^2+(H2O)12}.}
\label{u12}
\end{figure}

To investigate the solvation and spin-orbit effects on the lowest excitation energy of the uranyl ion in an aqueous solution, we performed calculations for the bare uranyl ion and the uranyl ion with one and two solvation shells, 
namely, \ce{UO2^2+(H2O)4} and \ce{UO2^2+(H2O)12}, 
using both the SFX2C-1e-EOM-CCSD and CD-based X2CAMF-EOM-CCSD methods. 
The molecular structures of these clusters are shown in Fig. \ref{u12}. 
The lowest excitation energies of \ce{UO2^2+}, \ce{UO2^2+(H2O)4}, and \ce{UO2^2+(H2O)12} are summarized in Table \ref{uo2sol}. 
We note that the calculation of \ce{UO2^2+(H2O)12} cluster correlated 120 electrons and 860 virtual spin orbitals, which is the most computationally intensive calculation reported in this paper.
A Cholesky threshold of $10^{-4}$ has been used for these calculations.

\begin{table}
\caption{The lowest excitation energies (in eV) of \ce{UO2^2+}, \ce{UO2^2+(H2O)4}, and \ce{UO2^2+(H2O)12} obtained from the SFX2C-1e- and X2CAMF-EOM-CCSD calculations. 
The SFX2C-1e excitation energies correspond to
those for the lowest triplet state.
The spin-orbit coupling corrections ($\Delta$SOC) were obtained as the differences between 
the X2CAMF and SFX2C-1e excitation energies. }
\begin{tabular}{cccc}
\hline \hline
 & \ce{UO2^2+} & \ce{UO2^2+(H2O)4}  & \ce{UO2^2+(H2O)12} \tabularnewline
\hline 
SFX2C-1e & 2.88 & 2.93  & 2.93 \tabularnewline
X2CAMF   & 2.56 & 2.57 & 2.56 \tabularnewline
\multirow{1}{*}{$\Delta$SOC} & -0.32 & -0.36  & -0.37 \tabularnewline
\hline \hline
\end{tabular}\label{uo2sol}
\end{table}

By comparing the SFX2C-1e excitation energies for \ce{UO2^2+}, \ce{UO2^2+(H2O)4}, and \ce{UO2^2+(H2O)12}, 
we observe that the solvation effects from the first and second solvation shells increase the excitation energy of 
the lowest triplet state by 0.05 and less than 0.01 eV, respectively. 
The corresponding corrections in the X2CAMF calculations with spin-orbit coupling are also rather insignificant, amounting to approximately 0.01 eV from the first solvation shell and 0.01 eV for the second solvation shell. 
The trend for the X2CAMF results is consistent with that for the lowest triplet state in the SFX2C-1e calculation.
To further investigate the dependence of solvation effects on the structures of the first solvation shell, we calculated the lowest excitation energy for \ce{UO2^2+(H2O)5}. 
In the scalar-relativistic SFX2C-1e calculations, the excitation energy 
for the lowest triplet state is 2.95 eV, while the lowest excitation energy in the X2CAMF calculation with spin-orbit effects included is 2.58 eV. 
The differences in the excitation energies of \ce{UO2^2+(H2O)4} and \ce{UO2^2+(H2O)5} are 
less than 0.02 eV. 
Based on the above results, we may hence safely conclude that solvation effects are insignificant for the lowest excitation energy of aqueous uranyl.

The differences between the results of the X2CAMF and the SFX2C-1e calculations define the spin-orbit corrections. 
Spin-orbit coupling reduces the lowest excitation energy of the bare uranyl ion by 0.32 eV. 
The spin-orbit corrections obtained from calculations with solvation shells are approximately 0.04 eV larger than those in the bare ion. 
The variation resulting from different solvation structures, i.e., less than 0.05 eV, is generally small in comparison to the overall corrections. 
For the uranyl ion, spin-orbit coupling hence has a more significant impact on the excitation energies than solvation effects.

\section{Summary and outlook}
\label{summarySec}
We present an implementation of Cholesky decomposition-based X2CAMF-CCSD(T) and EOM-CCSD methods.
By combining CD and AO-based algorithms, the present implementation eliminates the construction and storage of the $\langle ab||cd\rangle$- and $\langle ab||ci\rangle$-type integrals and intermediates. 
The CD-based implementation enables spinor-based relativistic CC calculations for medium-sized molecules containing heavy atoms while correlating more than 1000 spinors. 
Our benchmark calculations on ionization energies, bond-dissociation energies, and vertical excitation energies of uranium-containing small molecules demonstrate that a Cholesky threshold of $10^{-4}$ is sufficient to achieve a balanced trade-off between maintaining chemical accuracy and computational efficiency.
We further demonstrate the applicability of CD-based relativistic CC methods in calculations of the bond-dissociation energy of \ce{UF_6} and the lowest excitation energies of \ce{UO_2^2+} solvated with up to 12 water molecules.

The development of analytic gradients for the CD-based X2CAMF-CCSD(T) and EOM-CCSD methods is our priority for future work.
While the CD-based implementation for the X2CAMF-CC methods reported in this work 
effectively reduces disk and memory requirement,
the floating-point operation counts are still similar to those of the traditional implementations.
To further extend the applicability of X2CAMF-CCSD(T) and EOM-CCSD to larger molecules, it thus is necessary to use techniques 
that work with a compact representation of the correlation space in order to reduce the floating-point operations,
including methods based on
pair-natural orbitals \cite{meyerPNOCIStudiesElectron1973,neeseEfficientAccurateApproximations2009,riplingerEfficientLinearScaling2013} and/or local-correlation methods \cite{liLinearScalingLocal2002,fedorovCoupledclusterTheoryBased2005,katsLocalCC2Electronic2006}. 
In addition, leveraging the high-level parallelization provided by GPU-programming for the CD-based implementation could further accelerate X2CAMF-CCSD(T) and EOM-CCSD calculations. 
These techniques have the potential to render relativistic CC methods more accessible and useful in studying complex chemical systems containing heavy atoms.

\section*{Supporting Information}

The basis sets employed in the manuscript in CFOUR format. 
Raw data for the benchmark calculations shown in Figure \ref{bchmk}.

\begin{acknowledgement}
This work has been supported by the Department of Energy, Office and
Science, Office of Basic Energy Sciences under Award Number DE-SC0020317.
The computations at Johns Hopkins University were carried out at Advanced
Research Computing at Hopkins (ARCH) core facility (rockfish.jhu.edu),
which is supported by the National Science Foundation (NSF) under
grant number OAC-1920103. This work has been supported in Mainz and Saarbrücken by the
Deutsche Forschungsgemeinschaft (DFG) via project B05 of the TRR 146 "Multiscale Simulation Methods for Soft Matter Systems".
\end{acknowledgement}
\clearpage
\bibliography{references}
\clearpage
\begin{tocentry}
    \begin{center}
    \includegraphics[width=1.0\textwidth]{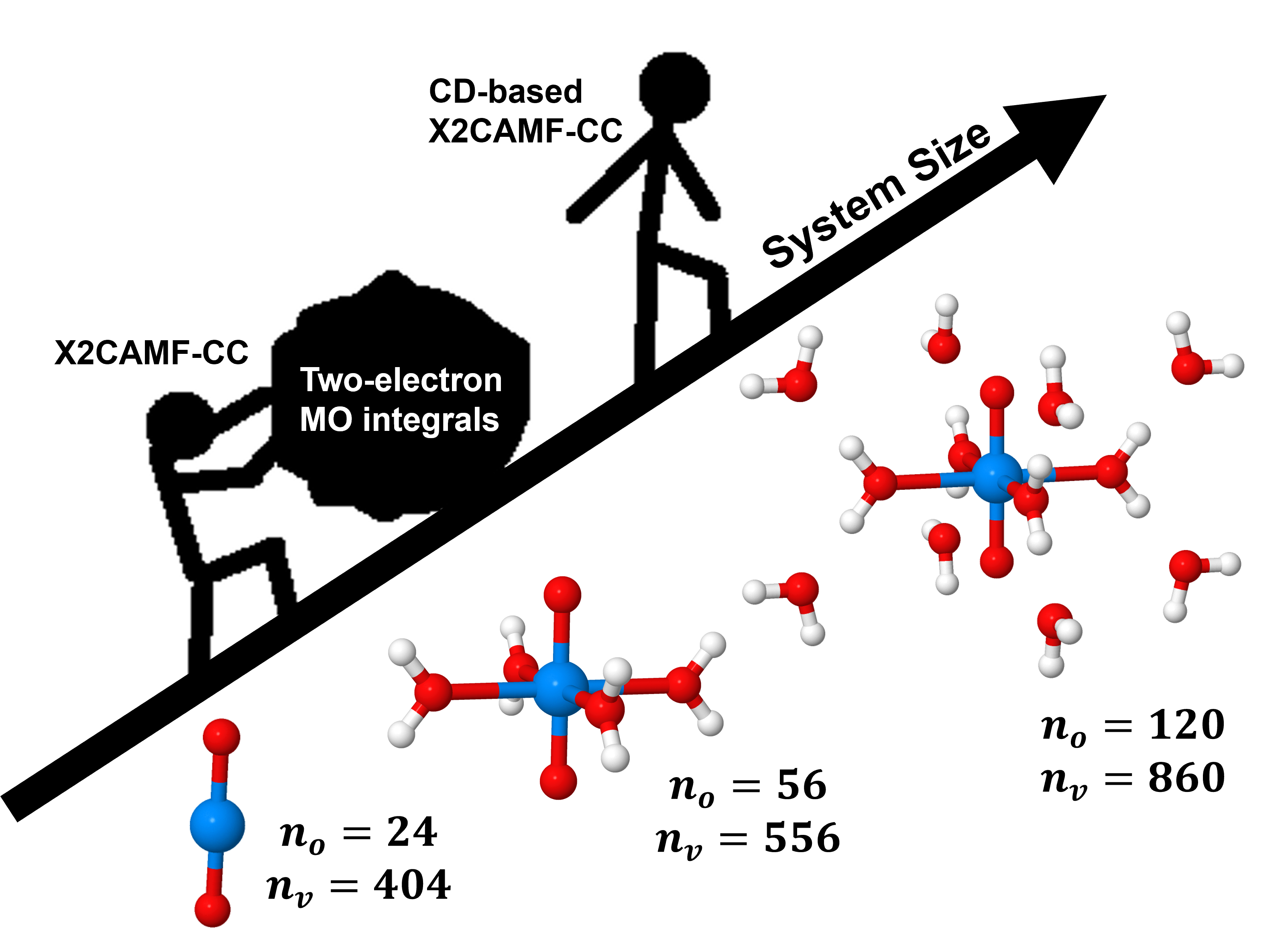}
    \end{center}
\end{tocentry}
\end{document}